\documentclass[%
reprint,
showpacs,preprintnumbers,
amsmath,amssymb,
aps,
prstab,
floatfix,
]{revtex4-1}

\usepackage{graphicx}
\usepackage{dcolumn}
\usepackage{bm}
\usepackage{paralist}
\usepackage{enumitem}
\usepackage{xcolor}
\usepackage{color}
\usepackage[colorlinks=true, citecolor=cyan, urlcolor=black, linkcolor=blue, pdfborder={0 0 0}]{hyperref}

\begin{document}

\preprint{Hughes-2017}

\title{On-Chip Laser Power Delivery System for Dielectric Laser Accelerators}

\author{Tyler W. Hughes}
 \email{twhughes@stanford.edu}
\author{Si Tan}%
 \email{stan1987@stanford.edu}
\author{Zhexin Zhao}
\author{Neil V. Sapra}
\author{Kenneth J. Leedle}
\author{Huiyang Deng}
\author{Yu Miao}
\author{Dylan S. Black}
\author{Olav Solgaard}
\author{James S. Harris}
\author{Jelena Vuckovic}
\author{Robert L. Byer}
\author{Shanhui Fan}
\affiliation{Stanford University, Stanford, CA 94305}
\author{Yun Jo Lee}
\author{Minghao Qi}
\affiliation{Purdue University, West Lafayette, IN 47907}
\collaboration{ACHIP Collaboration}
\noaffiliation{}

\date{\today}

\begin{abstract}
We propose an on-chip optical power delivery system for dielectric laser accelerators based on a fractal `tree-branch' dielectric waveguide network. This system replaces experimentally demanding free-space manipulations of the driving laser beam with chip-integrated techniques based on precise nano-fabrication, enabling access to orders of magnitude increases in the interaction length and total energy gain for these miniature accelerators. Based on computational modeling, in the relativistic regime, our laser delivery system is estimated to provide 21\,keV of energy gain over an acceleration length of 192\,$\mu$m with a single laser input, corresponding to a 108\,MV/m acceleration gradient. The system may achieve 1\,MeV of energy gain over a distance less than 1\,cm by sequentially illuminating 49 identical structures. These findings are verified by detailed numerical simulation and modeling of the subcomponents and we provide a discussion of the main constraints, challenges, and relevant parameters in regards to on-chip laser coupling for dielectric laser accelerators.
\end{abstract}

\maketitle


\section{\label{sec:intro} Introduction}


In recent years, dielectric laser accelerators (DLAs) have demonstrated acceleration gradients (energy gain per unit length) approaching 1\,GV/m \cite{peralta2013demonstration, wootton2016demonstration, leedle2015dielectric, leedle2015laser, breuer2013laser, plettner2006proposed, cesar2017nonlinear, kozak2017dielectric, kozak2017acceleration}, several orders of magnitude higher than those attainable by conventional linear accelerator systems based on microwave-driven metal waveguide structures \cite{solyak2009gradient}. This breakthrough is made possible by the advent of advanced nano-fabrication techniques \cite{ohtsu2012near, simakov2017diamond, arakawa2013silicon, lim2014review, thomson2016roadmap} combined with the fact that dielectric materials may sustain electric fields close to 10 GV/m when illuminated by ultra-fast NIR laser pulses \cite{stuart1995laser, tien1999short, jee1988laser}. High acceleration gradients may allow DLAs to accomplish significant energy gains in very short lengths, which would enable numerous opportunities in fields where compact and low-cost accelerators would be useful, such as medical imaging, radiation therapy, and industrial applications \cite{koyama2014parameter,england2014dielectric, wootton2016dielectric}.

Since DLA structures are already driven at their damage thresholds, apart from finding methods to increase material damage thresholds, achieving high total energy gain from DLA will fundamentally require extending the interaction length between the incoming laser pulse and the particle beam. Several proof of principle DLA experiments \cite{wootton2017recent,peralta2013demonstration} have demonstrated high acceleration gradients by use of free-space manipulation of the laser pulse, including lensing, pulse-front-tilting \cite{hebling1996derivation, akturk2004pulse}, or multiple driving lasers \cite{kenleedle, mcneur2016elements}. However, these techniques require extensive experimental effort to perform and the system is exceedingly sensitive to angular alignment, thermal fluctuations, and mechanical noise. An on-chip laser power delivery system would allow for orders of magnitude increases in the interaction lengths and energy gains achievable from DLA by replacing free-space manipulation with precise nano-fabrication techniques.

\begin{figure}[htbp!]
\includegraphics[width=\columnwidth]{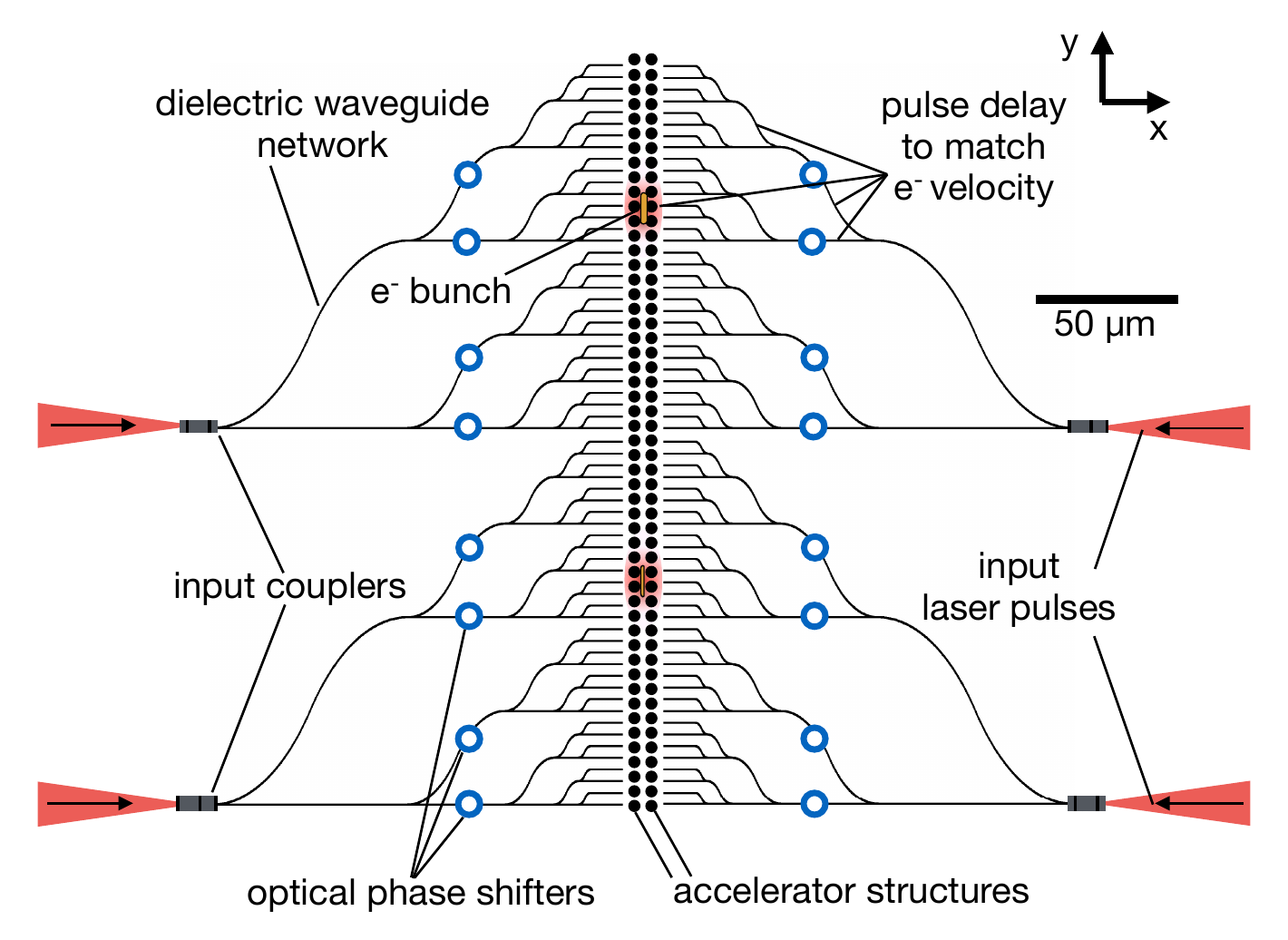}
\caption{\label{fig:struct} Two stages of the DLA laser coupling `tree-branch' structure.  The electron beam travels along the y-axis through the center of this structure.  The laser pulses are side coupled with optical power shown in red.  Black regions define the on-chip waveguide network.  Blue circles represent the optical phase shifters used to tune the phase of the laser pulse.}
\end{figure}

In designing any laser power delivery system for DLA, there are a few major requirements to consider. (1) The optical power spatial profile must have good overlap with the electron beam side profile. (2) The laser pulses must be appropriately delayed along the length of the accelerator to arrive at the same time as the moving electron bunches. (3) The optical fields along each section of the accelerator must be of the correct phase to avoid dephasing between the electrons and incoming laser fields. To accomplish all three of these requirements, we introduce a method for on-chip power delivery, which is based on a fractal `tree-branch' geometry introduced in Fig.~\ref{fig:struct}. In this paper, we provide a systematic study of the structure's operating principles, the optimal range of operating parameters, and the fundamental trade-offs that must be considered for any on-chip laser coupling strategy of the same class. Through detailed numerical modeling of this design, we estimate that the proposed structure may achieve 1\,MeV of energy gain over a distance less than 1\,cm by sequentially illuminating 49 identical structures.

The paper is organized as follows: in Sec.~\ref{sec:model}, we introduce the working principles and components of the proposed laser coupling system.  In Sec.~\ref{sec:constraints}, we overview the main constraints facing this system.  In Sec.~\ref{sec:param}, we present the findings of a parameter study investigating the structure.  In Sec.~\ref{sec:discussion}, we discuss the limitations and benefits of the proposed structure and propose future directions for this work before concluding in Sec.~\ref{sec:conclusion}.  The assumptions and values used in the parameter study are validated by discussion in the appendix sections.

\section{\label{sec:model}System Model}

We first introduce the proposed `tree branch' waveguide geometry, which is diagrammed in Fig.~\ref{fig:struct}.  The electron beam to be accelerated is propagating along the y-axis in the central accelerator gap.  We first couple the laser pulses to the on-chip dielectric waveguides by use of input couplers.  The optical power is then split a series of times and directed by waveguide bends to illuminate the entire length of the accelerator gap. Integrated phase shifters are used to tune the phase of each pulse upon exiting the waveguides and may be optimized for maximum acceleration. The accelerating structures are placed adjacent to the waveguide outputs. In this study, we choose to investigate silicon dual-pillar accelerator structures, similar to those used in \cite{leedle2015dielectric}. The entire device is mirrored over the center plane and is driven by laser inputs on either side.  Two stages of the structure are shown in Fig.~\ref{fig:struct}, although several more may be implemented in series, assuming availability of several phase-locked laser sources.  Electron beam focusing elements may be implemented between stages as needed.  A detailed overview of the individual components is given in Appendix~\ref{appx:components}.

A fractal waveguide geometry is chosen as it evenly illuminates the accelerator gap with minimal use of 50-50 splitters.  Furthermore, the waveguide bends are designed such that the laser pulse arrival at the accelerator gap is delayed to coincide with the arrival of the electron bunch as it propagates through the structure.  This requirement sets strict conditions on the bending radius required at each section.  The mathematical details are outlined in Appendix \ref{appx:treebranch}.

\section{\label{sec:constraints}Constraints}

In the analysis of our system, we consider four main factors that will ultimately limit the acceleration gradients and energy gains attainable by our structure.

\paragraph{Laser-induced damage of the DLA and waveguide materials.} To avoid damage of the structure, the electric fields in the system may never exceed the damage thresholds of the dielectrics used.  The laser damage threshold for dielectric materials is highly favorable at short pulse durations, with sustainable electric fields that scale roughly as $\tau^{-1/2}$ for $\tau >$ 1 ps and approach $\tau^{-1}$ scaling for fs pulses \cite{stuart1995laser,stuart1996nanosecond}.  Amongst the materials considered in this study, SiO$_2$ has the highest damage fluence threshold of 2.5\,J/cm$^{2}$ at 800 nm wavelength, followed by Si$_3$N$_4$ at 0.65\,J/cm$^{2}$ and Si at 0.18\,J/cm$^{2}$ \cite{soong2014particle}. For a 100\,fs pulse propagating in vacuum, these correspond to peak fields of 3.7, 7.0, and 13.7\,GV/m, respectively.  In Appendix \ref{appx:min-field} we derive approximations for the minimum input electric field allowable before damaging either our input coupler or accelerator structure.

\paragraph{Optical nonlinearities in the materials.} Optical nonlinear effects are encountered when propagating through the waveguides and may cause significant pulse distortion.  This may result in either damage or dramatic reduction of the acceleration gradient due to phase mismatching. Although a full treatment is given in Appendix~\ref{appx:nonlinear}, the most prominent nonlinear effects in our structure are self-phase modulation (SPM) and self-focusing (SF). For a pulse with a given peak power, the effects of SPM scale in proportion to the lengths of the waveguide sections. On the other hand, for a given pulse peak power, SF is made worse at longer pulse durations, and is less of a concern than SPM for the pulse durations that we are interested in (10\,fs$-$10\,ps). We derive approximations for the minimum input electric fields before these nonlinearities occur, which are described in Appendix~\ref{appx:min-field}.  These approximations were further shown to be consistent with the full treatment.
 
\paragraph{Power loss.}The tree-branch structure introduces several sources of power loss. (1) Input coupling loss, (2) splitting loss, (3) bending loss, and (4) waveguide scattering loss, which are discussed in detail in Appendix~\ref{appx:components}. In addition, since the optical power is split in half at each bend, the power of each output port will be reduced by at least a factor of $2^{N_s}$ with respect to the input facet, where $N_s$ is the number of splits. These effects mean that the damage will be more concentrated at the input facet for a larger number of splits, since the optical power will be highly attenuated by the time it reaches the output ports.  Waveguide power loss due to scattering must be considered for structures with interaction lengths greater than the cm scale \cite{yamada2011silicon}.  However, we neglect these effects in this study because we focus on shorter waveguide segments.

\paragraph{DLA structure resonance characteristics versus input pulse bandwidth.}The DLA structures are designed to resonantly enhance the optical fields. As derived in Appendix~\ref{appx:resonances}, the field enhancement is proportional to the square root of the quality factor of the DLA structures. This resonance is used to increase the acceleration gradient while avoiding damage at the input facet. However, if the pulse bandwidth is small with respect to the bandwidth of the accelerator, the pulse will not efficiently couple into the DLA structure.  These effects are modeled directly in the parameter study, following the treatment outlined in Appendix~\ref{appx:time-freq}.

\section{\label{sec:param}Parameter Study}
With the system components and constraints introduced, we now present a parameter study to understand the fundamental trade-offs and optimal working parameters of an on-chip optical power delivery system for DLA.  A software package~\cite{hughes2017github} was written to separately model each component and combine the results to generate an estimate for the acceleration gradient and energy gain assuming a set of parameters, which are outlined in Table~\ref{tab:params}.  The choice of these parameters is validated in Appendix~\ref{appx:components}.

\begin{table}[htb]
\caption{\label{tab:params} Parameters assumed in the study.}
\centering
\begin{tabular}{lccc}
\hline
Parameter & Symbol & Value & Units \\
\hline
Wavelength & $\lambda$ & 2 & $\mu$m \\
Electron speed / speed of light & $\beta$ & 1 & - \\
DLA periods per waveguide & M & 3 & - \\
Input coupler efficiency & $\eta_c$ & 0.6 & - \\
Splitting efficiency & $\eta_s$ & 0.95 & - \\
Bending efficiency & $\eta_b$ & 0.95 & - \\
Accelerating gradient at Q = 1 & $G_{Q=1}$ & 0.0357 & $E_0$ \\
Input coupler - first split length & $L_0$ & 10 & $\mu$m \\
DLA pillar radius & $R_{pillar}$ & 981 & nm \\
\hline
\end{tabular}
\end{table}

The four constraints are modeled using the approximations introduced in the previous section.  To model the DLA structures, we use the two-dimensional finite-difference frequency-domain method (FDFD) \cite{shin2012choice} to simulate a waveguide feeding into the dual pillar structures.  The pillars are assumed to have infinite extent out of the plane, neglecting fringing effects. A Lorentzian fit to the frequency response of the DLA structures is used to estimate the Q-factor of our structure.  Using this Q-factor value, the corresponding acceleration gradient, and the scaling discussion in Appendix~\ref{appx:resonances}, we may estimate the acceleration gradient at any Q-factor.  The phase at each output is assumed to be corrected for maximum acceleration.

\begin{figure}[htpb!]
\includegraphics[width=\columnwidth]{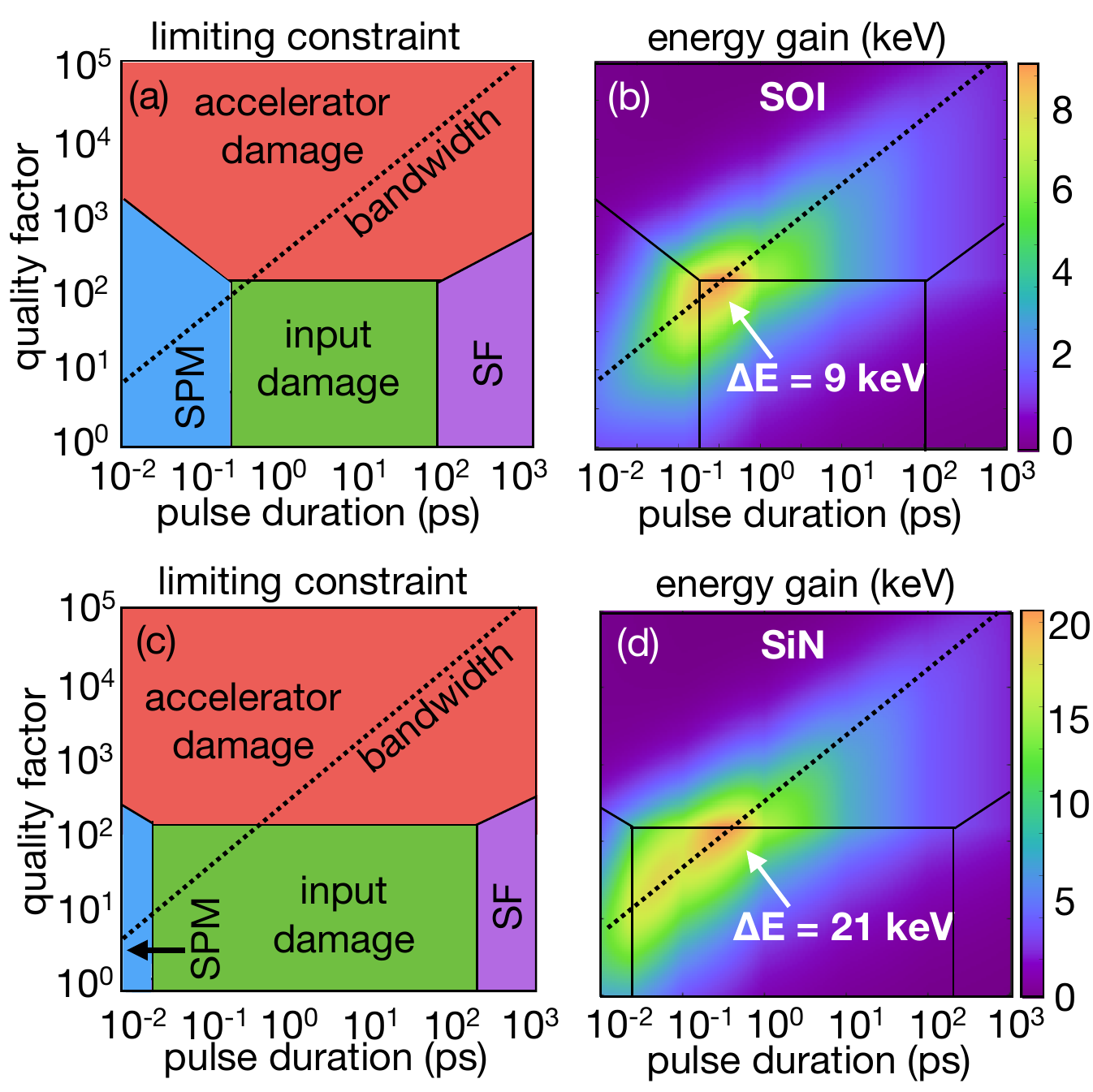}
\caption{\label{fig:param1} Results from the parameter study. A single stage of the tree branch structure is considered, with interaction length of 192 $\mu$m, corresponding to $5$ power splits and $2^5 = 32$ output ports. In (a-b), Silicon-on-Insulator (SOI) waveguides are assumed. in (c-d), a Si$_3$N$_4$/SiO$_2$ waveguides are assumed. For each Q-factor and pulse duration, we compute the maximum input field achievable before damage or nonlinearity occurs. The limiting constraint is shown in (a) and (c).  The maximum Q-factor before the pulse bandwidth exceeds the DLA resonator bandwidth is given by the dotted line. The energy gain from one section is plotted in (b) and (d). The acceleration gradient follows the same trends as the energy gain.  Optimal operating regimes are clearly visible from the plots.}
\end{figure}

We first choose to examine a single stage with interaction length of 192 $\mu$m, corresponding to 5 splits and 32 output ports.  This number is chosen as it gives a reasonable balance between acceleration gradient and energy gain.  Over a range of pulse durations ($\tau$) and Q-factors ($Q$), we first compute the minimum peak electric field at input that will cause either damage or nonlinear pulse distortion using the expressions in Appendix~\ref{appx:resonances}.  Then, for relativistic electrons, we use the assumed parameters to compute the achievable acceleration gradient and energy gain from this section.  In Fig.~\ref{fig:param1}, we show the limiting constraints for each $\tau$ and $Q$, as well as the energy gain from a single stage for waveguides with core materials of Si and Si$_3$N$_4$.

From Fig.~\ref{fig:param1}, we see that, for a given geometry, there is an optimal combination of $\tau$ and $Q$ where the energy gains and acceleration gradients are maximized. For a structure with an interaction of length 192 $\rm{\mu m}$, this point at $\tau$ = 341 (322) fs and $Q$ = 157 (154) for waveguide cores made of Si (Si$_3$N$_4$).  A full list of the results are displayed in Table~\ref{tab:results}.  Using a SiN waveguide system, we may expect to achieve 1 MeV of energy gain at 109 MV/m gradients by running 49 stages in series.  

There are several competing effects that lead to the existence of this optimal point. First, for a given pulse peak power, shorter pulse durations will generally lead to higher acceleration gradients because the materials will exhibit higher electric field damage thresholds. However, this effect is limited by the occurrence of SPM at a certain input field. Furthermore, if the pulse is too short with respect to the Q-factor of the DLA structures, the pulse will not couple efficiently to the accelerator gap due to the pulse bandwidth being smaller than the structural bandwidth. Secondly, higher Q-factors lead to resonantly enhanced fields inside of the DLA structure and higher acceleration gradients as a result \cite{deng2017design}. However, if the Q-factor is too high, these enhanced fields will cause the accelerator structures to damage.


\begin{table}[htb]
\caption{\label{tab:results} Optimal results from the parameter study, for waveguides fabricated from SOI and SiN.}
\centering
\begin{tabular}{lccc}
\hline
Metric & Value & Value & Units \\
 & (SOI) & (SiN) & \\
\hline
Acceleration gradient & 45.3 & 107.5 & MV/m \\
Energy gain per stage & 8.7 & 20.6 & keV \\
Input peak electric field & 1.0 & 2.4 & GV/m \\ 
Pulse duration & 341 & 322 & fs \\
Q-factor & 156.71 & 154.0 & - \\
Pulse energy & 0.36 & 11.3 & nJ \\
Number of stages for 1\,MeV & 116 & 49 & - \\
Stage length & 192 & 192 & $\mu$m \\
Waveguide core width & 0.78 & 2 & $\mu$m \\
Waveguide core height & 220 & 400 & nm \\
\hline
\end{tabular}
\end{table}

\begin{figure}[htpb!]
\includegraphics[width=\columnwidth]{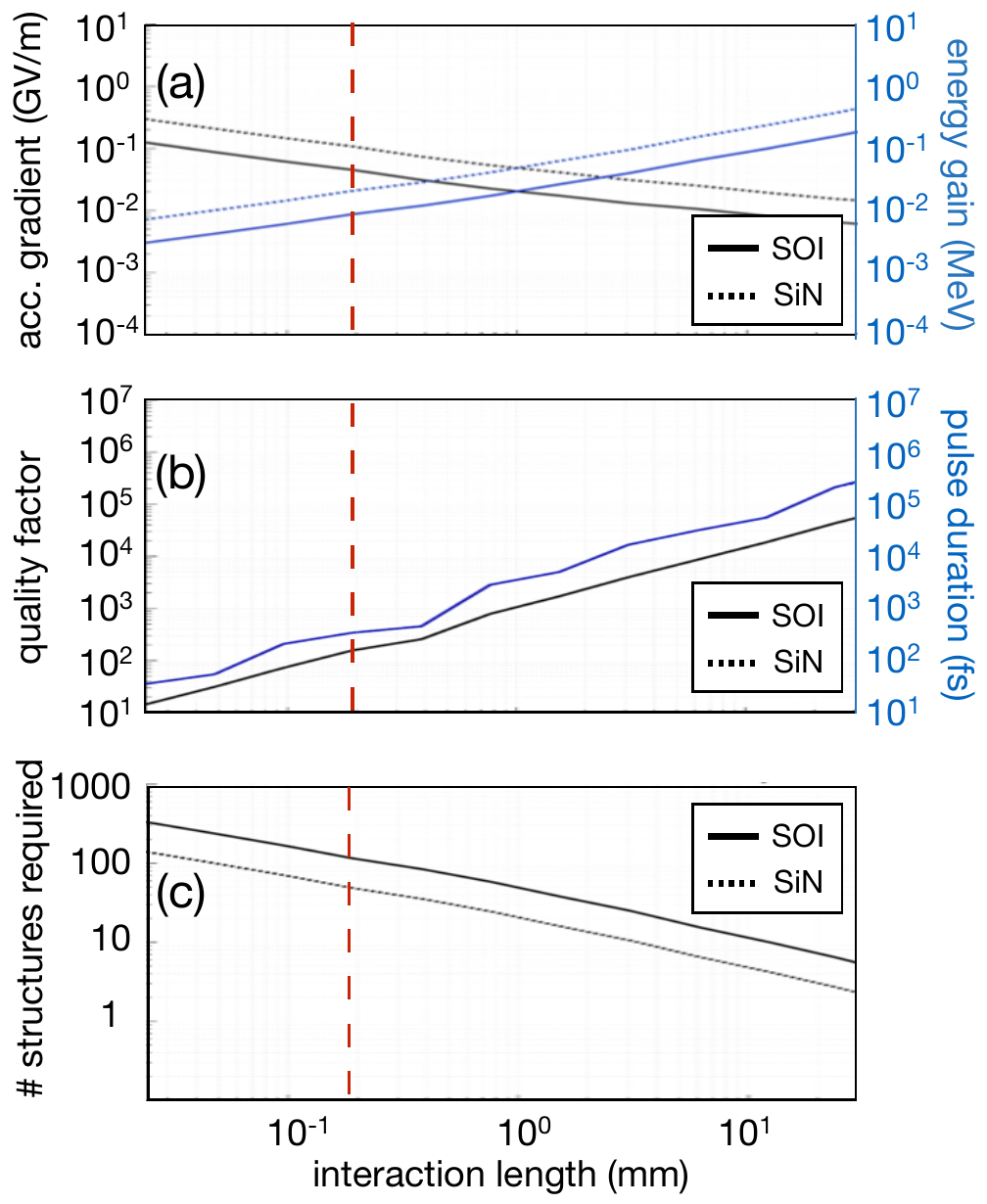}
\caption{\label{fig:param2}  Scaling of optimal parameters as a function of the interaction length.  The red dotted line corresponds to an interaction length of 192 $\mu$m, which is the length used in Fig. \ref{fig:param1}.  a) The optimal energy gains and acceleration gradients achievable from this structure as a function of interaction length for both SOI and SiN structures.  b) The optimal set of pulse duration and Q-factor corresponding to the highest energy gain and acceleration gradient at each interaction length.  The curves for SOI and SiN are overlaid.  c) The number of stages of each interaction length required to reach 1 MeV of total energy gain.}
\end{figure}

To investigate how these results depend on the interaction length, we run several of these simulations over a range of structures with different numbers of splits, keeping track of the optimal $\tau$, $Q$, acceleration gradient, and energy gain of each structure.  The results are presented in Fig.~\ref{fig:param2}.

From Fig.~\ref{fig:param2}a, we note that as the interaction lengths become longer, the achievable acceleration gradients decrease due to the increased losses introduced by the greater number of splits, combined with the increased nonlinearities and concentration of optical power at the input facet. On the other hand, the energy gain increases with greater interaction length until 10 cm. Thus, there is an intrinsic trade-off between having a high acceleration gradient and a large energy gain per laser input. Therefore, the choice in interaction length should be determined by the acceleration gradients and energy gains required by the application. For instances where high acceleration gradient is preferred, a smaller interaction length per laser is optimal, meaning less splits. However, for applications where high total energy gain is a more important figure of merit, it may be beneficial to use a coupling structure with many splits and long interaction length, but lower acceleration gradient.  These metrics will also depend on the availability of several phase-locked laser sources and the experimental difficulties associated with coupling them to several input couplers.

From inspecting Fig.~\ref{fig:param2}b, we see that the optimal $\tau$ and $Q$ increases as the structure becomes larger. Thus, the longer the interaction length we wish to supply with this tree-branch structure, the more resonance we require in the DLA structure. For a longer interaction length, more splits must be performed, which puts additional burden on the input facet relative to the DLA structure. This, in turn, requires greater resonant enhancement at the accelerator gap to offset, and a subsequently larger $\tau$ to match the structural bandwidth.


\section{\label{sec:discussion}Discussion}

We now discuss some strategies for improving on the results presented in this parameter study, as well as potential challenges and future directions.

\begin{figure}[htpb!]
\includegraphics[width=\columnwidth]{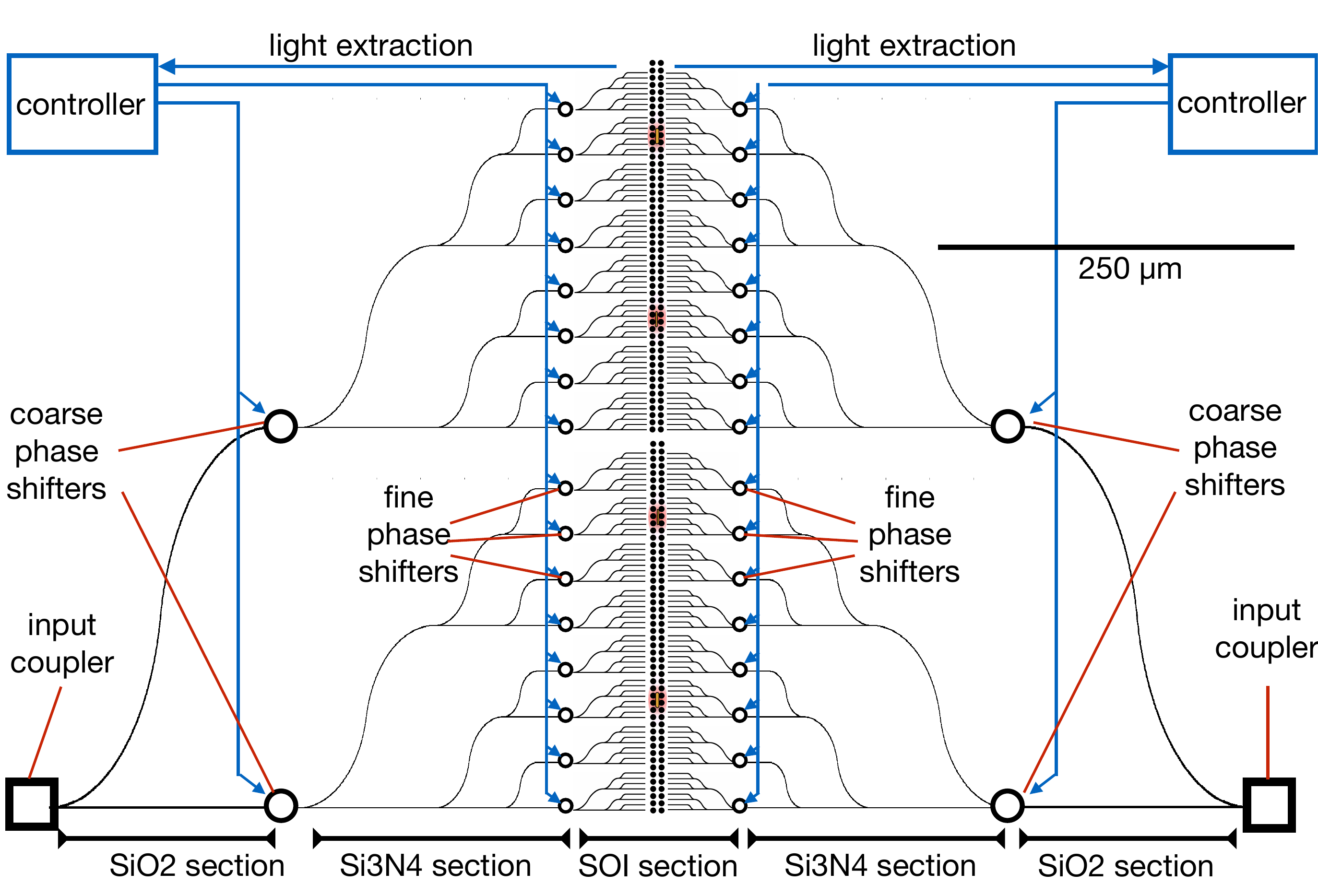}
\caption{\label{fig:PhC} Schematic of a hybrid structure for DLA laser coupling. Center: an SOI tree branch-DLA geometry optimized for tight bends and compact waveguides. This is fed by a Si$_3$N$_4$/SiO$_2$ waveguide section with relatively higher damage threshold, and lower nonlinearities. This section is then fed by an all SiO$_2$ power delivery section as described in the discussion section. Coarse and fine phase shifters are used in different splitting sections. In general, high power handling and low nonlinearity materials are used towards the input end, whereas tight bending and fine phase control structures are used towards the DLA end of the structure.}
\end{figure}

We notice that SiN waveguide systems may supply much higher acceleration gradients than those of SOI systems.  This is due to the favorable damage and nonlinear properties of Si$_3$N$_4$ compared to Si.  However, as shown in Appendix~\ref{appx:components}, SiN waveguides have high bending loss at bend radii below 50~$\mu$m due of the low refractive index of Si$_3$N$_4$ compared to Si.  Therefore, to mitigate the effects of damage and nonlinearities in our waveguide system while maintaining adequate bending radii required for pulse delay, one solution is to implement a hybrid system comprising of a laser power delivery system optimized for high power handling to feed a series of smaller tree-branch structures optimized for tight bends. A diagram of this setup is given in Fig \ref{fig:PhC}. 

Waveguiding systems for this high power handling region may be based on hollow-core photonic crystals, high damage threshold materials, such as silica or silicon nitride, or weakly guided waveguide modes. The section closer to the DLA could then be implemented in SOI allowing for tight bending radii, compact waveguide networks, and fine phase control.  The DLA structures may also be integrated directly on the same chip as the inner power delivery system.  Multiple of these hybrid systems may be driven in series, each with an individual driving laser. This would require multiple phase-locked mode-locked fiber lasers, approaches to which have been demonstrated in \cite{rauschenberger2002control, cundiff2003colloquium, jones2000carrier, shelton2001synchronization, washburn2004phase}.  The relative merits of large interaction length power delivery systems vs. multiple driving lasers will depend on their respective engineering challenges, such as chip-to-chip coupling \cite{shoji2002low}, alignment and stability of input coupling multiple lasers, and availability of these sources.  

One set of attractive options for further improving the acceleration gradients and energy gains achievable with an on-chip waveguide power delivery system involve engineering the group velocity dispersion (GVD) of the waveguides.  One possible strategy involves pre-chirping the input pulse to compensate for the GVD.  Then, the optical power may be initially spread in the temporal domain, mitigating high damage bottlenecks near the input facet.  Later, with the presence of GVD, the structure may be designed such that the pulse re-compresses at the accelerator structure.  Additionally, we may use GVD to balance out SPM effects in our waveguides.  With the proper amount of GVD, a temporal soliton may be formed for a given power, which will propagate without distortion, potentially allowing for higher operating powers and acceleration gradients. A similar technique was recently demonstrated to compensate for the SPM effects in short DLA structures \cite{cesar2017nonlinear}. These are promising avenues for exploration, but were not considered in this work with the intention of establishing a conservative baseline for the merits of on-chip laser coupling.

Based on the presented geometry, there is a clear need for resonant DLA structures to enhance the fields at the accelerator gap. For the parameters discussed, the optimal Q-factors were shown to be around 150. Previous work on optimizing DLA structures for high acceleration gradient has shown that periodic dielectric mirrors may be useful in raising quality factors and field enhancement in DLA structures \cite{hughes2017method, mizrahi2004optical, niedermayer2017designing, wei2017dual}. However, achieving DLA structures with these Q-factors may be difficult with current fabrication tolerances. Furthermore, even slight deformation due to both electron collision with the DLA structure and the presence of high power optical pulses would degrade the Q-factors of fabricated structures. Therefore, experimental verification is required to determine whether such resonant structures can survive operation in a DLA. If it is not the case, then alternative schemes must be presented to eliminate the need for resonance. 

The next stage of this study will involve experimentally verifying the parameters assumed, including the waveguide damage thresholds, input coupling loss, splitting loss, bending loss, and acceleration gradients.  Then, a proof of principle optical test will be performed on a simple system before acceleration experiments with electron beams are performed.

\section{\label{sec:conclusion}Conclusion}
We have presented a method for accomplishing chip-based laser power delivery for DLA applications. For a stage length of 192 $\mu$m, our method predicts acceleration gradients greater than 100\,MV/m, and 1\,MeV of energy gain in less than 1\,cm with 49 structures integrated in series. Our proposal has a major advantage over free-space laser coupling techniques in that it is arbitrarily scalable in interaction length and total energy gain. This is of critical importance in enabling DLA to move from proof-of-principle to application stage, where large energy gains are a critical figure of merit. 

\begin{acknowledgments}
We wish to acknowledge everyone in the ACHIP collaboration.  Special thanks to Kent Wootton and R. Joel England for their guidance and comments.  This work was supported by the Gordon and Betty Moore Foundation (GBMF4744).
\end{acknowledgments}

\appendix

\section{\label{appx:components}Structure Components}
To validate the assumptions made in the previous study, we will now discuss the individual components involved in the on-chip laser coupling system.

\subsection{Input Coupling}
The proposed structure first requires a strategy to couple light from the pump laser to the on-chip accelerating structure. In general, couplers must have (1) high coupling efficiency, (2) bandwidth large enough to couple entire pulse spectrum, (3) high power handling and minimized hot spots. Input coupling may be accomplished by use of end coupling, focusing the laser beam directly onto the waveguide cross section, or vertical coupling schemes, such as grating couplers. End coupling can achieve insertion losses as low as $0.66\;\text{dB (85.9\%)}$ over a bandwidth of roughly 10 THz \cite{pu2010ultra}, but is cumbersome to perform experimentally for a large number of inputs and constrains the input and output coupling ports to be located on the edges of the chip. Vertical couplers provide the benefit of relative flexibility in alignment and positioning on chip. The coupling efficiency of these devices varies drastically depending on the complexity of the grating coupler design, from an efficiency of $>30\%$ to $>90\%$ \cite{taillaert2006grating}. However, highly efficient broadband couplers capable of sustaining large bandwidths still provide design challenges, with the state-of-the-art fully-etched structures able to provide $67\%$ coupling efficiency with a 3 dB bandwidth of 60 nm at 1550 nm \cite{ding2013ultrahigh}. 
In this study, we assume a coupling power efficiency of $60\%$ with a substantially wide bandwidth to accommodate that of our pulse (up to $\approx$ 10 THz for a 50 fs pulse), which is reasonably achievable with end coupling. Additional investigation into design of ultra-broadband vertical couplers must be considered in order to guarentee coupling of the femtosecond pulsed lasers used in this experiment.

\subsection{Waveguides}
\begin{figure}[hbtp!]
\includegraphics[width=\columnwidth]{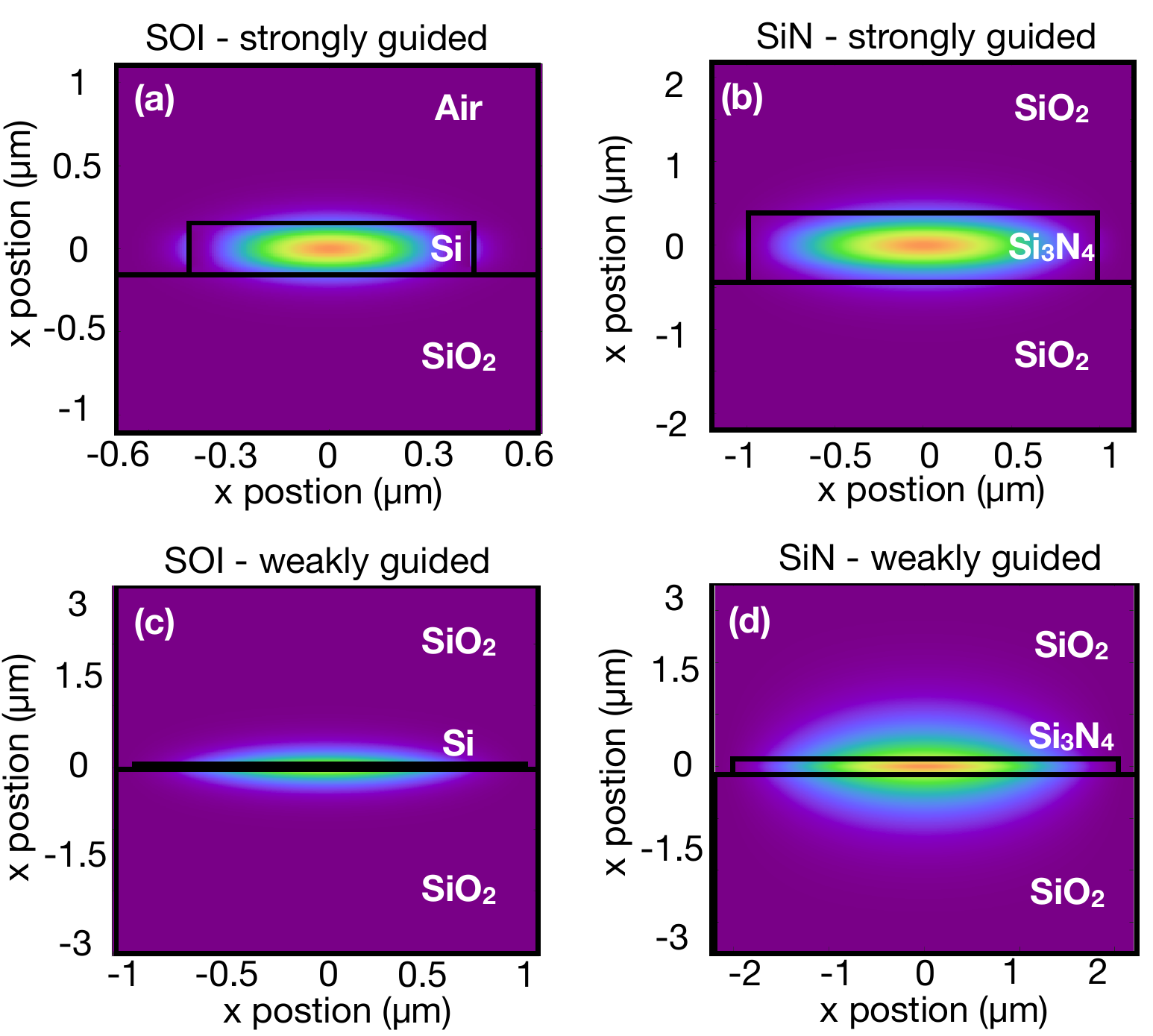}
\caption{\label{fig:waveguides} Waveguide geometries and corresponding horizontal electric field components \cite{fallahkhair2008vector}. (a-b) Strongly confined modes. (c-d) Weakly confined modes. (a) and (c) are SOI material platforms whereas (b) and (d) are Si$_3$N$_4$/SiO$_2$ materials. Waveguide core heights in (a-d) are given by 220, 400, 60, and 100 nm, respectively. Waveguide core widths are given by 0.78, 1.6, 2, and 4 $\mu$m, respectively.}
\end{figure}

Waveguides are a critical component of laser coupling. Schematics of the waveguide systems and their field distributions are shown in Fig. \ref{fig:waveguides}. We have explored two general classes of wave-guiding systems: (1) tightly confined systems and (2) weakly confined systems. Weakly confined waveguide modes have a small difference between mode effective index and cladding index, which results in the optical power being spread over a larger area and into the cladding material, which generally has preferable damage and nonlinearity properties. However, as we will show in the next section, our simulations show that weakly confined modes, with $n_{eff} - n_{core}$ of about 0.1, have almost 0$\%$ power transmission for bend radii less than 10 $\mu$m. In our tree-branch structure, we  require bend radii on this order of magnitude to achieve the required pulse delay to matching to the electron bunch, therefore weakly guided waveguides were not considered for the particular tree branch structure in this parameter study.

Additionally, we have explored material systems of SOI and Si$_3$N$_4$/SiO$_2$ due to their common use as waveguide core materials. SOI-based waveguides would be simpler to integrate with the silicon DLA structure and electron gun. Additionally, there exists a much larger body of previous work on fabrication of silicon material systems for applications such as phase control, especially in the LIDAR community \cite{yaacobi2014integrated, kwong20111}. However, Si$_3$N$_4$/SiO$_2$ waveguide have favorable nonlinear and damage properties when compared to SOI.

\subsection{Splitters}
After the initial input coupling step, splitters are used to distribute the laser power across the DLA structure, which will further contribute to insertion loss. Y-junctions, while theoretically lossless, are limited in efficiency due to the finite resolution of fabrication processes. Experimental characterization of such devices indicate losses on the order of $1\;\text{dB}$ \cite{zhang2013compact}. Recent advances in topology optimization techniques have allowed for new designs with much higher efficiencies. Using Particle Swarm Optimization, devices have been produced with theoretical insertion losses of $0.13\;\text{dB}$ and an experimentally determined value of $0.28 \pm 0.02\;\text{dB}$ \cite{zhang2013compact}. As even more sophisticated techniques of optimization have been developed, the insertion loss of simulated designs has reached $0.07\;\text{dB}$ \cite{lalau2013adjoint}. This method of using adjoint optimization has been further expanded to enforce fabrication constraints on the permitted designs, thus allowing one to expect greater agreement between simulated and fabricated structures \cite{piggott2017fabrication}. As a consequence of the rapid progress made in this field and the efforts to ensure robustness of device to fabrication tolerance, we have used an insertion loss per splitter of $0.22\;\text{dB}$, or $95\%$ efficiency, for the parameter study.

\subsection{Bends}

The bending radius is uniquely chosen to give enough extra propagation distance to provide a delay of the pulse between different output ports, which is matched to the electron velocity. We derive conditions on the radius of curvature required for each bend for the particular tree branch structure in Appendix \ref{appx:treebranch}. The required radius depends on the electron velocity ($\beta c_0$) and group index of the waveguide mode ($n_g$), and generally becomes smaller as the waveguides approach the DLA structure. Assuming the tree-branch geometry used in this work, following the derivation in Appendix \ref{appx:treebranch}, there is a condition on the group index of the waveguide system that may achieve the required delay given an electron speed
\begin{equation}
n_g\beta \geq 1.
\end{equation}
Thus, for sub-relativistic electrons ($\beta < 1$), higher index materials are required for the waveguides. For example, for a $\beta$ of $1/3$, a group index of $n_g > 3$ is required, which may not be satisfied by a standard SiN waveguide geometry. Thus, in sub-relativistic regimes, SOI waveguides are the optimal choice.

\begin{figure}[hbtp!]
\includegraphics[width=\columnwidth]{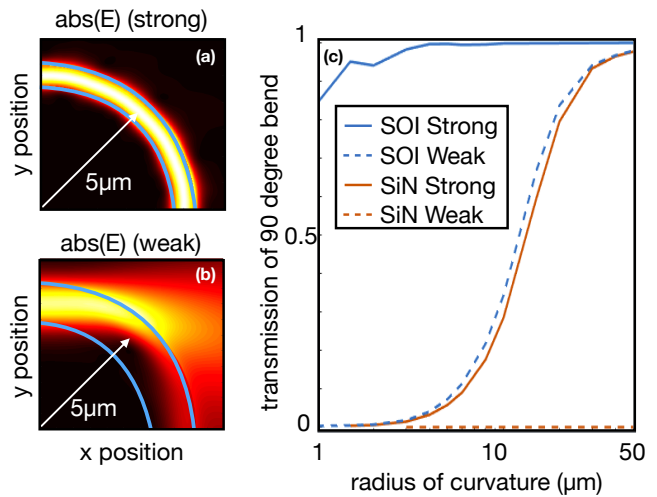}
\caption{\label{fig:bend}(a). Absolute value of E field for strongly guiding SOI waveguide. (b). Absolute value of E field for weakly guiding SOI waveguide. (c). Comparison of bending loss as a function of bend radius for the 4 waveguides from Fig. \ref{fig:waveguides}.} 
\end{figure}

In Fig.~\ref{fig:bend} we show the optical power transmission through a series of bends and waveguide geometries using Finite Difference Frequency Domain (FDFD) \cite{shin2012choice} and an established effective two-dimensional approximation to the three-dimensional structure \cite{smotrova2005cold}. For tightly confined SOI waveguide modes, the bending radius can reach as low as 2 $\mu$m before there is significant loss. However, for weakly confined SOI modes and strongly confined SiN modes, the power transmission is less than 50$\%$ until the radius exceeds 20\,$\mu$m. For our purposes, this kind of bending loss is unacceptable as radii on the order of $10$\,$\mu$m are required close to the DLA structure to perfectly match the electron velocity. However, if we relax the delay requirement in favor of larger bend radii, we may still use strongly confined SiN modes. Based on a calculation following appendix \ref{appx:treebranch}, if we wish to keep all SiN waveguides above 40\,$\mu$m radius of curvature, we will experience a 25\,fs mismatch in peak pulse arrival to electron arrival. For a pulse duration of 250\,fs, this will have negligible effect on the acceleration gradient. Therefore, in our parameter study, we assume strongly confined waveguide modes and bends that are large enough to achieve transmission of 95\%. Many of these issues may be reconciled by choosing a hybrid waveguide system, as shown in Fig.~\ref{fig:PhC}, in which different materials and waveguide modes are used at different distances from the central DLA structure. For instance, while Si waveguides may be optimal close to the DLA, where tight bends are a requirement, it may make sense to use either SiN waveguides or weakly confined modes towards the input end of the structure, where damage and nonlinearities are more important considerations. We did not consider these options directly in our parameter study.

\subsection{Phase Shifters}
Phase shifters are an essential component in the DLA system for ensuring proper phase matching between the electrons and photons. While it is simple to do phase tuning in free-space for a single stage DLA with macroscopic delay stages, waveguide integrated phase shifters for long interaction or multi-stage DLAs will be experimentally complicated. 
In order to achieve a sizable energy gain and gradient over a given interaction length, a high level of precision and stability in the phase of each section is required. Through a Monte Carlo simulation of the output phase of each waveguide, we found that, for an interaction length of 1 mm, we require phase stability and precision of about 1/100 of a radian (16\% of a cycle) to achieve sustained energy gain. 

There are a few strategies to implement integrated phase shifters, (1) Thermal/thermal-optic effect \cite{kwong20111,kwong2014chip}, (2) Electro-optic effect, (3) Mechanical techniques, such as piezo controlled elements \cite{poot2014broadband}. For this application, we will require a full $2\pi$ range of phase control of each output port with a resolution of 1/100 of a radian, and a modulation bandwidth of $\sim$ 1 kHz to correct for environmental perturbations. 

Rather than supplying each waveguide output port with a phase shifter with these properties, it may be possible to have dedicated `fine' and `coarse' phase shifters as we move through the splitting structure. Furthermore, some degree of relative fixed phase between output ports may be accomplished by precision fabrication. 

\begin{figure}
\includegraphics[width=\columnwidth]{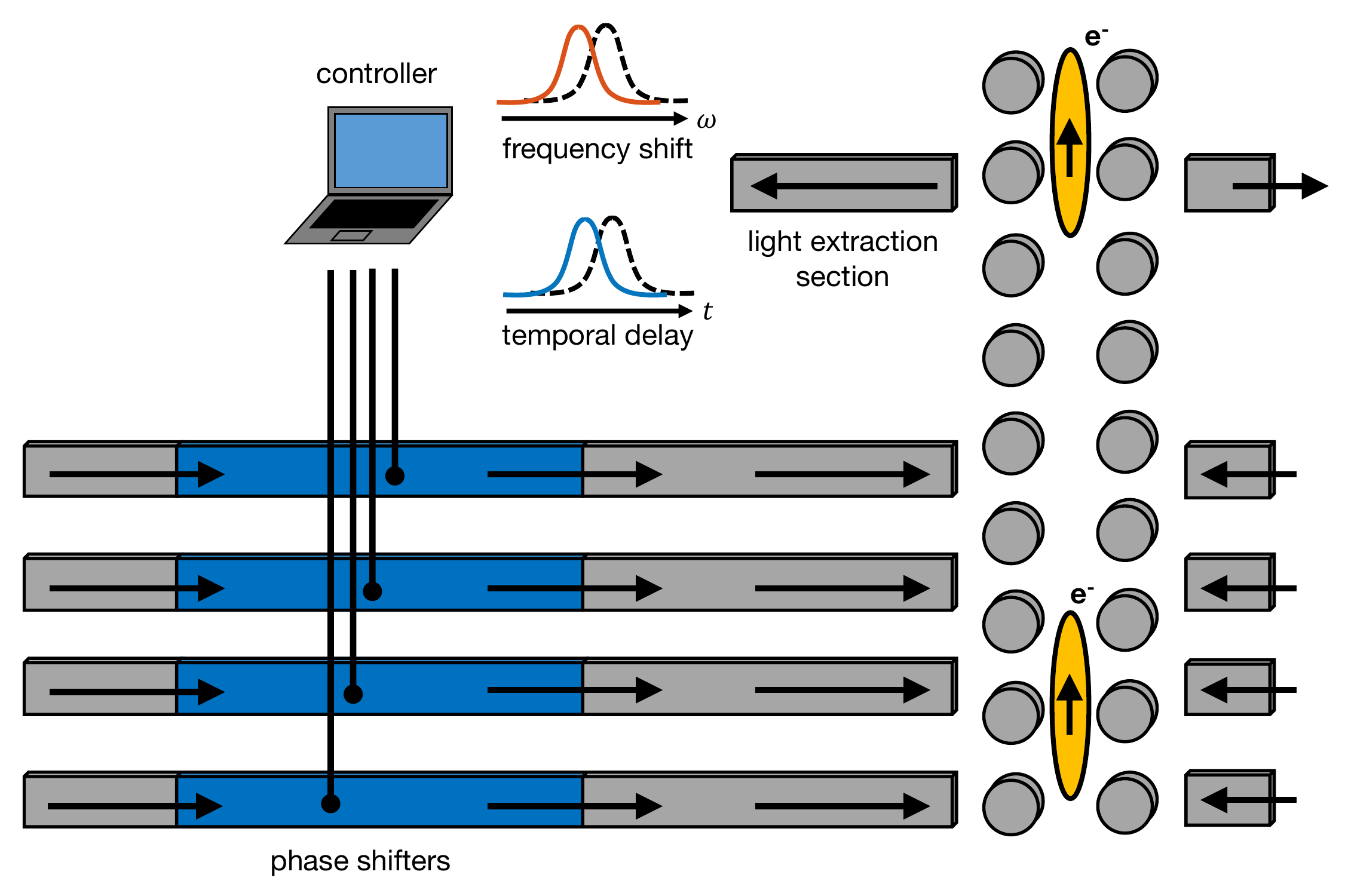}
\caption{\label{fig:phase} Feedback system for automatic phase control. A dedicated light extraction section is added to the accelerator. Light is radiated from the electron beam transversing the DLA structures and the frequency content and/or timing of the light is sent to a controller. The phase shifts of each waveguide are optimized with respect to either the frequency or the delay of the signal. After several runs, the system should converge to stable operation.}

\end{figure}

To further mitigate the challenges associated with operating these multiple phase shifters during acceleration, we may implement a feedback control loop, which is described in Fig. \ref{fig:phase}. 


\subsection{DLA Structures\label{sec:DLA}}

\begin{figure}[htbp!]
\includegraphics[width=\columnwidth]{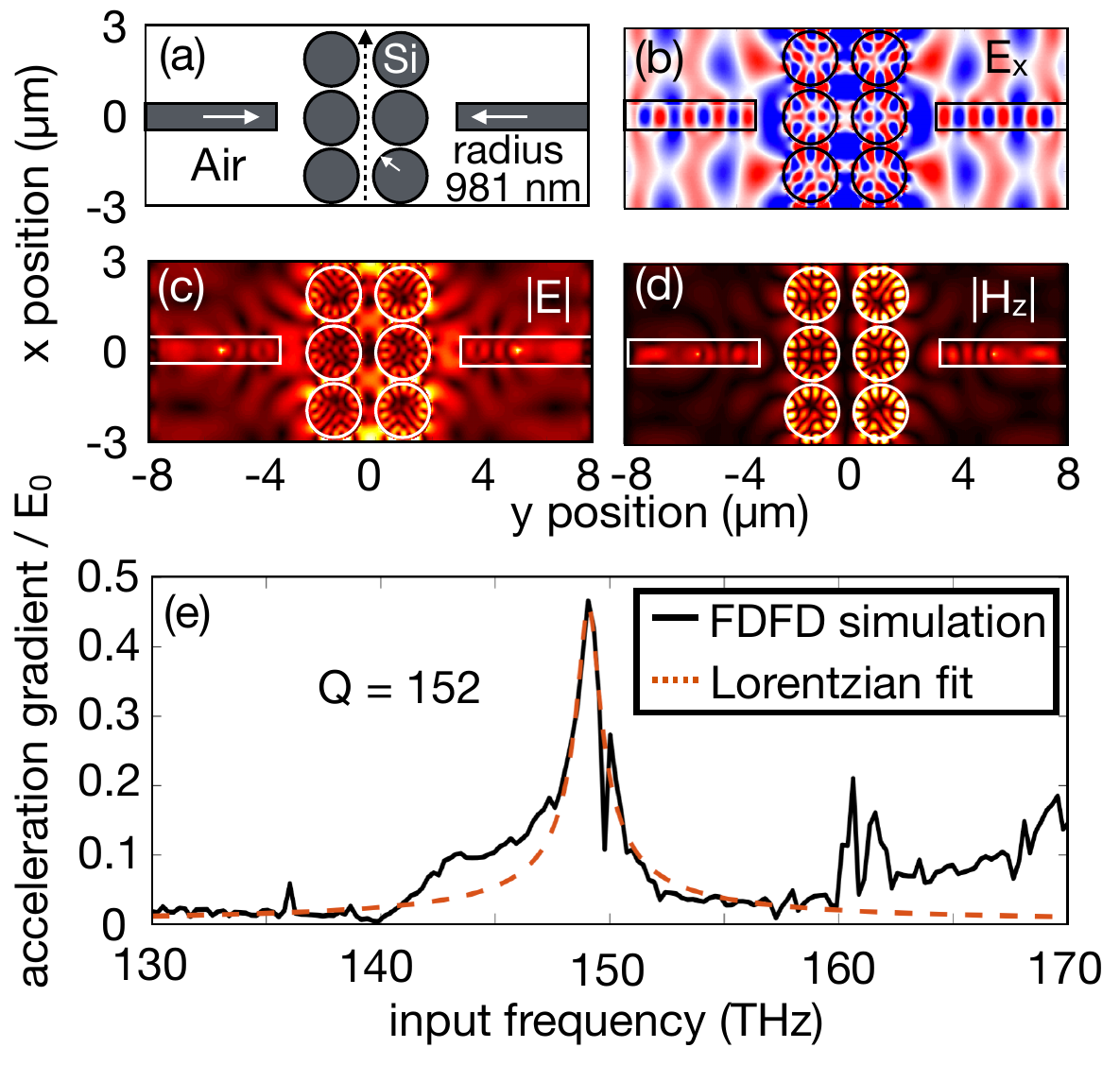}
\caption{\label{fig:DLA} a) A schematic of the waveguide to DLA connection. Silicon dual pillars of optimized radius of 981 $\mu$m are used. The distance between the waveguide end facet and the pillars may also be optimized to give resonant effects. b) The accelerating electric field during one time step. c) absolute value of the electric field. d) Absolute value of the transverse magnetic field. Resonant enhancement in the dual pillars is clearly visible. d) Absolute value of the acceleration gradient as a function of frequency, normalized by the peak electric field in the waveguide. Computed numerically with FDFD for the two-dimensional structure in (a-d). The waveguide refractive index was approximated using \cite{smotrova2005cold}. A Lorentzian line shape is fit to the square of this plot. The square root of this fit is shown in red. Based on the Lorentzian fit, a Q-factor of 161 was determined.}
\end{figure}

We assume silicon dual-pillar DLA structures in the parameter study, but the choice is arbitrary and can be changed to other materials or designs depending on the optical power delivery system. Fig. \ref{fig:DLA}a shows a schematic of the setup, along with the results of a frequency scan of the structure and the Lorentzian fit.

\begin{figure}[htbp!]
\includegraphics[width=\columnwidth]{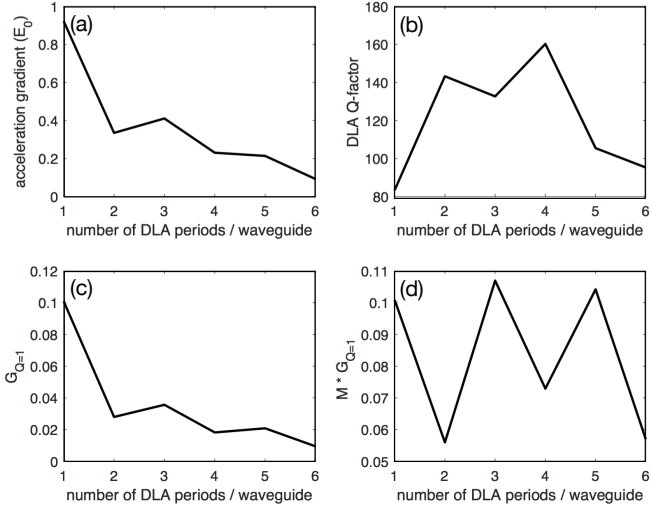}
\caption{\label{fig:M_scan} Investigation into changing the number of periods of DLA powered by a single waveguide.  All DLA structures are assumed to be Si dual pillar as in Fig. \ref{fig:DLA}.  a) Acceleration gradient as a function of number of periods per waveguide, normalized by the peak field in the waveguide ($E_0$). b) Quality factor of the DLA structures after having radius optimized for maximum acceleration gradient. c) The acceleration gradient corresponding to Q = 1, as derived in Appendix \ref{appx:resonances}. d) The acceleration gradient at Q = 1 times the number of periods (M), which gives an estimate to the relative energy gain from each waveguide.  Since this plot is relatively constant, it appears that supplying more periods per waveguide does not increase the energy gain supplied by each waveguide.  Therefore, tighter waveguide networks are preferred.}
\end{figure}

Coupling from waveguides to DLA structures may be done by optimizing both the spacing between these two elements and the  DLA geometry parameters, such as pillar radius. For an optimized structure, back reflection may be minimized. It will be of great importance in future experiments to integrate the waveguide system and the DLA structure on the same chip. Thus, the height of the pillar structure may be constrained to be equal to that of the waveguide core and 500 nm thick SOI platforms may be a good starting point for testing these integrated systems. One waveguide is able to serve multiple DLA periods, a discussion is shown in Fig. \ref{fig:M_scan} for dual pillar structures. The findings suggest that adding more periods of DLA does not increase total energy gain from a single waveguide, thus it may be best to have as small a spacing between waveguides as possible.

\section{\label{appx:treebranch} Tree-branch structure: Velocity matching to electron beam}

\begin{figure}
\includegraphics[width=0.8\columnwidth]{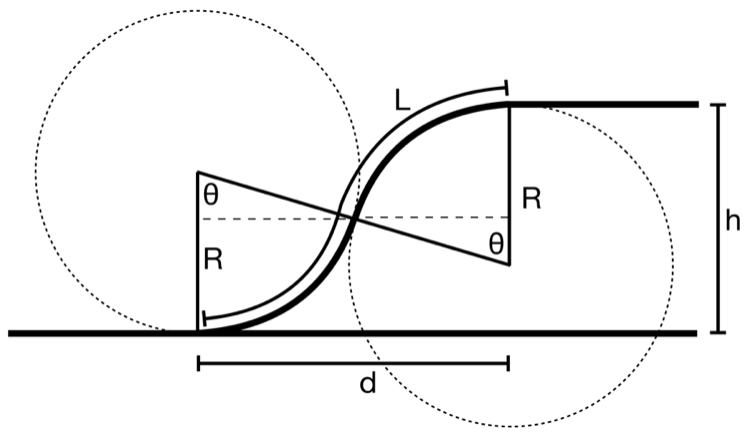}
\caption{\label{fig:app_bend} Diagram of a single bend in the tree-branch structure with an optical pulse incident from the left. The bend has radius $R$, accomplishes a vertical climb of $h$ over a horizontal distance $d$. The total length of the bend section is $L$. The electron travels from bottom to top in this configuration. We wish to find an $R$ such that an optical pulse traveling through the bent section is delayed by the same amount of time for the electron to travel the vertical distance $h$}
\end{figure}

Using the circular bending geometry as described in Fig.~\ref{fig:app_bend}, we may provide a delay to the pulse to match the electron velocity in the DLA structure. For a given vertical distance $h$ and waveguide group index $n_g$, we seek to set a condition on an $R$ to to accomplish this. First, we may establish the value of the bend angle `$\theta$' as
\begin{equation}
	\theta = \Bigg\{
 \begin{array}{ll}
  \cos^{-1}(1-h/2R) &\text{if } h < 2R \\
  \pi/2   &\text{if } h \geq 2R
 \end{array}.
\end{equation}
When $h \geq 2R$, we use two 90-degree bends and extend the intermittent length with a vertical waveguide section.
From this, we can express the horizontal distance $d$ as
\begin{equation}
	d = 2R\sin(\theta),
\end{equation}
and the total length of the bent waveguide as
\begin{equation}
	L = \Bigg\{
 \begin{array}{ll}
  2R\theta    &\text{if } h < 2R \\
  h + (\pi-2)R    &\text{if } h \geq 2R
 \end{array}.
\end{equation}

To now set a condition on $R$, we insist that the pulse timing delay between the curved waveguide and the straight waveguide is equal the time needed for the electron to travel a distance $h$. The difference in length between the curved waveguide and straight waveguide is simply $L-d$, thus the timing delay of the pulse is given by
\begin{align}
 \Delta t_{pulse} &= \frac{n_g}{c_0}(L-d)\\
 	 &= \frac{n_g}{c_0}\Bigg\{
 \begin{array}{ll}
  2R(\theta-\sin(\theta))   &\text{if } h < 2R \\
  (h + R(\pi-4))     &\text{if } h \geq 2R
 \end{array}.
\end{align}
The electron has a velocity of $\beta c_0$, so it's timing delay is given by 
\begin{equation}
\Delta t_{e^-} = \frac{h}{\beta c_0}
\end{equation}
Setting these two equal and solving for `R', we find that
\begin{equation}
	R = \frac{h}{\beta n_g}\Bigg\{
 \begin{array}{ll}
  2(\theta-\sin(\theta))^{-1}    &\text{if } h < 2R \\
  \frac{\beta n_g - 1}{4-\pi}   &\text{if } h \geq 2R
 \end{array}.
\end{equation}
Thus, for extended interaction lengths where $h >> 2R$, we require that $\beta n_g > 1$ for a positive (and physical) solution for $R$. Equivalently, for low $\beta$, we require large $n_g$ in order to sufficiently delay the pulse in order to match the low electron velocity.

\section{\label{appx:min-field}Derivation of minimum input field before damage or nonlinearities}

In this section we seek to give expressions for the maximum peak electric fields, denoted by $E_0$, that we may inject into our waveguide system before each constraint becomes relevant.

\textbf{(1) Input damage}: Fields at the input will be damaged if they exceed the damage threshold of the coupling material. Thus, we enforce that
\begin{equation}
E_0 < E_d(\tau).
\end{equation}

\textbf{(2) Accelerator damage}: With a given tree-branch structure, we introduce a total of $N_s$ separate 1$\to$2 power splits for an input pulse. Furthermore, we introduce some optical power loss characterized by the power efficiencies of the input coupler ($\eta_{c}$), splitters ($\eta_s$), and bends ($\eta_s$). Thus, the field at the output port of the laser coupling structure, $E_{out}$, is given by
\begin{equation}
E_{out} = E_0 \Big( 2^{-N_s} \eta_c \eta_s^{N_s} \eta_b^{N_s} \Big)^{1/2}.
\end{equation}

As we show in Appendix~\ref{appx:resonances}, resonance in the DLA structures with quality factor $Q$ will lead to a field enhancement in the accelerator gap that scales as $\sqrt{Q}$. Since our damage will be caused by the maximum field in the DLA materials, we assume there is another constant factor, $f_m$, relating the maximum field in the DLA material to the average field in the accelerator gap. From simulations, we estimate the value of $f_m$ to be 2. Thus, the maximum field in the DLA material is
\begin{align}
E_{mat} &= E_{out} f_m \sqrt{Q} \\
 &= E_0 f_m \sqrt{Q} \Big( 2^{-N_s} \eta_c \eta_s^{N_s} \eta_b^{N_s} \Big)^{1/2}.
\end{align}

We require the maximum field in the DLA material to be lower than the damage threshold, giving the constraint that
\begin{equation}
E_0 < E_d(\tau) \frac{2^{N_s/2}}{f_m \sqrt{Q}} \Big( \eta_c \eta_s^{N_s} \eta_b^{N_s} \Big)^{-1/2}.
\end{equation}

\textbf{(3) Self-phase modulation}: For a wave of power $P_0$ and wavelength $\lambda$ traveling a distance $L$ in a material with cross sectional area $A$, and nonlinear refractive index $n_2$, the accumulated SPM phase is given by \cite{teich1991fundamentals}
\begin{equation}
\Delta\phi_{SPM} = 2\pi \frac{n_2 PL}{A\lambda}.
\end{equation}

Since the optical power in our waveguides have optical power traveling in several materials, each with a different nonlinear refractive index, we define an effective $n_2$ for modeling that is given by
\begin{equation}
n_2^{(\textrm{eff})} = \frac{1}{P^{(tot)}}\sum_{j=1}^{num. mat.} n_2^{(j)} P^{(j)},
\end{equation}
where $P^{(tot)}$ is the total optical power carried by the waveguide and $P^{(j)}$ is the amount of power traveling in material `j'.

Furthermore, the optical power is being split in half at each bend, so we must take this into account in our SPM calculation. Taking into account the losses in our system, the final expression for the amount of SPM phase is 
\begin{equation}
	\Delta\phi_{SPM} = 2\pi \frac{n_2^{(\textrm{eff})} P_0 \eta_c}{A_{\textrm{eff}}\lambda}\sum_{i=0}^{N_s}\frac{\eta_s^i\eta_b^i L_i}{2^i}.
\end{equation}

Once the SPM phase reaches a value of $2\pi$, we notice pulse deformation leading to degradation of the acceleration gradient. This is confirmed by full simulations with our NLSE solver as described in Appendix~\ref{appx:nonlinear}. Thus, the constraint on our input field to avoid SPM effects is given by 
\begin{equation}
	E_0 < \Big(\frac{2\lambda }{n_2^{(\textrm{eff})} n c_0 \epsilon_0 \eta_c}\sum_{i=0}^{N_s} \frac{2^i}{\eta_s^i\eta_b^i L_i} \Big)^{1/2}.
\end{equation}

\textbf{(4) Self-focusing:}
The intensity-dependent change in refractive index may cause a lensing effect that can lead to self-focusing effects. In a guided mode this corresponds to a significant shrinking of the mode area and will lead to subsequent damage. The condition to avoid self-focusing is given by restricting the value of the `B integral' \cite{perry1994self} to $< \pi$, which is given by
\begin{equation}
B = 2\pi\frac{n_2}{\lambda}\int dz\ I(z) < \pi.
\end{equation}
Here $I(z)$ is the optical intensity over the propagation distance. We assume that the pulse is fully contained in the input waveguide section between the input coupler and the first split. This gives a conservative estimate of the self-focusing effect as it does not take into account power loss from splitting. This integral can be computed directly assuming a Gaussian pulse with FWHM pulse duration $\tau$ traveling in a waveguide with group index of $n_g$. 

The result of the integral is
\begin{equation}
B = 2\pi\frac{n_2^{(\textrm{eff})}P_0 c_0 \tau}{\lambda A_{\textrm{eff}} n_g \eta_c}\sqrt{\frac{\pi}{4
\ln(2)}} < \pi.
\end{equation}

Thus, this sets an additional constraint on the value of $E_0$ to avoid self-focusing, given by
\begin{equation}
E_0 < \Big(\frac{\lambda n_g}{c_0^2\tau n_2^{(\textrm{eff})} n \epsilon_0\eta_c}\sqrt{\frac{\pi}{4
\ln(2)}}\Big)^{1/2},
\end{equation}

\textbf{Summary}:
The maximum input fields that are safe to use before encountering each constraint are, thus, given by:
\begin{align}
E_0^{(inp.)} &= E_d(\tau) \\
E_0^{(acc.)} &= E_d(\tau) \frac{2^{N_s/2}}{f_m \sqrt{Q}} \Big( \eta_c \eta_s^{N_s} \eta_b^{N_s} \Big)^{-1/2}\\
	E_0^{(SPM)} &= \Big(\frac{2\lambda}{n_2^{(\textrm{eff})} n c_0 \epsilon_0 \eta_c}\sum_{i=0}^{N_s} \frac{2^i}{\eta_s^i\eta_b^i L_i} \Big)^{1/2}.\\
E_0^{(SF)} &= \Big(\frac{2 \lambda n_g}{2c_0^2\tau n_2^{(\textrm{eff})} n \epsilon_0\eta_c}\sqrt{\frac{\pi}{4
\ln(2)}}\Big)^{1/2}
\end{align}

\section{\label{appx:nonlinear} Nonlinearities}
To study waveguide nonlinearity, we solve a version of the nonlinear Schr\"{o}dinger equation (NLSE), which is typically used for describing nonlinear propagation of a pulse of duration between 10\,fs and 10\,ns. In this particular treatment, the solution for the electric field is assumed to be of form in Eq.~\ref{eq:field}, where the slowly varying envelope approximation and separation of variables of the modal distribution $F(x,y)$ and envelope $A(z,t)$ are used \cite{agrawal2007nonlinear}.
\begin{equation}
\mathbf{E(r},t)=\frac{\hat{x}}{2}\{F(x,y)A(z,t)\exp[i(\beta_0 z-\omega_0t)+\rm{c.c.}]\},
\label{eq:field}
\end{equation}
where $x,y$ are the transverse directions, $z$ is the propagation direction, $\beta_0$ is the propagation constant and $\omega_0$ is the optical frequency. The slowly varying envelop $A(z,t)$ obeys the form of the NLSE given in Eq. \ref{eq:NLS}, which can be solved by the split-step method \cite{weideman1986split}.
\begin{align}
&\frac{\partial A}{\partial z}+\frac{\alpha}{2}A+\frac{i\beta_2}{2}\frac{\partial^2A}{\partial T^2}-\frac{\beta_3}{6}\frac{\partial^3 A}{\partial T^3}\\
&=i\gamma \left(|A|^2A+\frac{i}{\omega_0}\frac{\partial}{\partial T}(|A|^2A)-T_{\rm{R}}A\frac{\partial |A|^2}{\partial T}\right),
\label{eq:NLS}
\end{align}
where $\gamma=2\pi n_2/(\lambda A_{\rm{eff}})$ is the nonlinear parameter per unit length and power and $A_{\rm{eff}}$ is the effective modal area. On the left hand side of this equation, the loss is incorporated into the second term with $\alpha$ being the loss of the waveguide in units of $\rm{m}^{-1}$. The 3rd and 4th terms indicate second and third order dispersion, with $\beta_2$ and $\beta_3$ being the respective dispersion coefficients. On the right hand side of the equation, 1st term is SPM, the 2nd term is self-steepening, and the 3rd term is Raman scattering.

For our proposed structure, the overall length of the waveguide is short ($\ll1$ m), hence material loss $\alpha$ can be neglected. The dispersion terms come from both the material dispersion and waveguide dispersion. These terms, $\beta_{2,\rm{wg}}$ and $\beta_{3,\rm{wg}}$, can be obtained from numerically solving for effective refractive index as a function of wavelength $n_{\rm{eff}}(\lambda)$, and are explicitly given as
\begin{align}
\beta_{2,\rm{wg}}&=\frac{\lambda^3}{2\pi c^2}\frac{d^2n_{\rm{eff}}}{d\lambda^2}\:,\\
\beta_{3,\rm{wg}}&=-\frac{3\lambda^4}{4\pi^2 c^3}\frac{d^2n_{\rm{eff}}}{d\lambda^2}-\frac{\lambda^5}{4\pi^2 c}\frac{d^3n_{\rm{eff}}}{d\lambda^3}.
\end{align}

We note that each term in Eq. \ref{eq:NLS} can be turned on/off to investigate its contribution, and we find that for this particular case, SPM is the dominant contribution to the nonlinearity, as turning on/off other terms does not yield a significant difference to the results. Hence, our choice of using SPM as the dominant nonlinearity in the parameter study is justified.

\section{\label{appx:resonances}DLA resonances}




In this part, we verify that the field enhancement inside the accelerator gap is approximately proportional to $\sqrt{Q}$. We approximate the dual pillar accelerator by a one dimensional Fabry-Perot cavity with resonant angular frequency $\omega_0$, quality factor $Q$ and cavity length $l_c$. Suppose the incident field and circulating field inside the cavity have amplitudes $E_{in}$ and $E_c$ respectively. At resonance frequency, the power dissipated from the cavity equals the incident power, which is $1/(2\eta_0)|E_{in}(\omega_0)|^2$ where $\eta_0$ represents vacuum impedance, and energy stored in the cavity is $1/(2\eta_0)|E_c(\omega_0)|^2\cdot 2l_c/c$. The quality factor can be expressed as:
\begin{equation}
Q = \omega_0 \frac{|E_c(\omega_0)|^2\cdot 2l_c/c}{|E_{in}(\omega_0)|^2}.
\end{equation}
Also, the cavity spectrum has a Lorentzian shape. So, the circulating and incident field amplitudes have the following relation,
\begin{equation}
\label{eq:Hc}
E_c(\omega) = \sqrt{\frac{Q}{2\pi \frac{2 l_c}{\lambda}}}e^{i\phi_0}\frac{\frac{\omega}{2Q}}{\frac{\omega}{2Q} -i(\omega - \omega_0)} E_{in}(\omega),
\end{equation}
where $2\pi/\lambda = \omega_0/c$ and $\phi_0$ is the phase difference between $E_c(\omega_0)$ and $E_{in}(\omega_0)$. Based on Eq. \ref{eq:Hc}, we can define the transfer function $H_c(\omega)$, such that $E_c(\omega) = H_c(\omega) E_{in}(\omega)$. We may further approximate $H_c$ by a gaussian function with the same peak and full width at half maximum and with the same phase near central frequency.
\begin{equation}
\label{eq:approx}
\begin{split}
H_c(\omega) & = H_c(\omega_0) \frac{1}{\sqrt{1 + [\frac{\omega - \omega_0}{\omega_0/(2Q)}]^2}} \exp \Big [i\arctan{\frac{\omega - \omega_0}{\omega_0/(2Q)}} \Big] \\
	& \simeq H_c(\omega_0) \exp \Big [ -2\ln(2) \big (\frac{\omega - \omega_0}{\omega_0/Q} \big)^2+ i\frac{\omega - \omega_0}{\omega_0/(2Q)} \Big ],
\end{split}
\end{equation}
where $H_c(\omega_0) = \sqrt{\frac{Q}{4\pi l_c /\lambda}} \exp(i\phi_0)$.

Assume the incident wave is a gaussian pulse with central angular frequency $\omega_0$, duration $\tau$ and peaked at $t = 0$, i.e. $E_{in}(t) = E_{in}(0)\exp(-2\ln(2) \frac{t^2}{\tau^2}-i\omega_0 t)$. In the frequency domain, it can be expressed as
\begin{equation}
\label{eq:Ein}
E_{in}(\omega) = E_{in}(0) \sqrt{\frac{\pi}{2\ln(2)}}\tau \exp \Big [- \frac{(\omega - \omega_0)^2 \tau^2}{8 \ln(2)} \Big ].
\end{equation}
From equations \ref{eq:Hc} - \ref{eq:Ein}, we can obtain the circulating field amplitude inside the accelerator gap in the frequency domain.
\begin{equation}
\label{eq:Ecfreq}
\begin{split}
E_c(\omega) = & E_{in}(0) \sqrt{\frac{\pi}{2\ln(2)}}\tau H_c(\omega_0) \exp \Big[ i \frac{\omega - \omega_0}{\omega_0/(2Q)} \Big ] \\
 	& \times \exp \Big \{ -2\ln(2)(\omega - \omega_0)^2 \Big[ \Big(\frac{Q}{\omega_0}\Big)^2 + \Big(\frac{\tau}{4 \ln(2)} \Big)^2 \Big] \Big \}
\end{split}
\end{equation}
After Fourier transform, the circulating field amplitude in time domain is
\begin{equation}
\label{eq:Ectime}
\begin{split}
E_c(t) = & E_{in}(0) \sqrt{\frac{Q}{2\pi \frac{2 l_c}{\lambda}}} e^{i\phi_0} \frac{\tau}{\sqrt{ \tau^2 + (\frac{4 \ln(2) Q}{\omega_0})^2}} \\
	& \times \exp \Big[ -2\ln(2) \frac{(t - \frac{2Q}{\omega_0})^2}{\tau^2 + (\frac{4 \ln(2) Q}{\omega_0})^2} -i\omega_0 t \Big].
\end{split}
\end{equation}

\section{\label{appx:time-freq}Acceleration gradient: time domain to frequency domain conversion}
Here we describe the correspondence between the time domain description of the acceleration gradient and the frequency domain approach that is used in this work and others \cite{hughes2017method}. We assume an input pulse $E_0(t)$, which leads to the creation of an accelerating field in the gap of $E_x(x,t)$ through the convolution with the corresponding impulse response function $h(x,t)$. In the frequency domain, this is done via multiplication of the pulse spectrum $E_0(\omega)$ with the transfer function $H(x,\omega)$
\begin{align}
 E_x(x,t) &= E_0(t) \ast h(x,t)\\
 E_x(x,\omega) &= E_0(\omega) H(x,\omega).
\end{align}

In the time domain, the acceleration gradient is expressed as an integral over the accelerating electric field over the particle's trajectory.
\begin{equation}
G = \frac{1}{L}\int_{0}^{L}dx\ E_x(x, t(x))
\end{equation}

If the electron moves uniformly in $\hat{x}$ with speed $\beta c_0$, then $x(t) = x_0 + \beta c_0 t$ and we may express the acceleration gradient as a function of the starting time, $t_0$, as
\begin{align}
G(t_0) &= \frac{1}{L}\int_{0}^{L} dx\ E_x(x,t_0 + x/\beta c_0) \\
 &= \frac{1}{L}\int_{0}^{L} dx\ \int_{-\infty}^{\infty}dt\ E_x(x,t)\delta(t-t_0- x/\beta c_0).
 \label{eq:t0_integral}
\end{align}

In previous works, such as Ref.~\cite{plettner2006proposed}, the acceleration gradient is computed by first performing a Finite Difference Time Domain (FDTD) simulation to record $E_x(x,t)$ along the gap for a series of time, and then maximizing the integral in Eq.~\ref{eq:t0_integral} with respect to $t_0$. However, we may equivalently do the computation in the frequency domain by Fourier transforming this equation with respect to $t_0$, which yields
\begin{align}
	G(\omega) &= \frac{1}{L}\int_{0}^{L} dx\ \int_{-\infty}^{\infty}dt\ E_x(x,t) e^{i\omega t - x/\beta c_0} \\
  &= \frac{1}{L}\int_{0}^{L} dx\ e^{-i\omega x/\beta c_0} \ \ \int_{-\infty}^{\infty}dt\ E_x(x,t) e^{i\omega t} \\
  &= \frac{1}{L}\int_{0}^{L} dx\ e^{-i\omega x/\beta c_0}\ E_0(\omega) H(x,\omega)\\
  &\equiv g(\omega)E_0(\omega)
\end{align}
Here $g(\omega)$ is the gradient normalized by the incident electric field at that frequency, $E_0(\omega)$, which is also described in Appendix~\ref{appx:length}. 
Now, by performing a series of FDFD simulations at discrete frequencies, we may estimate $H(x,\omega)$. Then, using the known pulse amplitude spectrum and phase information in $E_0(\omega)$, we can compute $G(\omega)$ as described. Finally, $G(t_0)$ can be determined by applying a inverse discrete Fourier transform on $G(\omega)$, and the acceleration gradient can then be found by taking the maximum of the absolute value of this quantity. Explicitly,
\begin{equation}
	G = \max_{t_0}|\mathcal{F}\{ g(\omega) E_0(\omega)\}|.
\end{equation}

\section{\label{appx:length}Finite length DLA structure bandwidth}

Let us assume that we have a DLA interaction length of $L$ along $\hat{x}$ with an incident laser pulse of the form $E_0(t)$ with spectrum $E_0(\omega)$. The laser is assumed to be uniform along the entire interaction length.

In the time domain, the accelerating fields along the acceleration gap may be expressed as the convolution of the input pulse with the gap's impulse response function, $h(x,t)$. In the frequency domain, these fields can be expressed as the multiplication of the pulse spectrum with the corresponding transfer function, $H(x,\omega)$.
\begin{align}
	E_x(x,t) &= E_0(t) \ast h(x,t)\\
 E_x(x,\omega) &= E_0(\omega) H(x,\omega)
\end{align}

The DLA structure is further assumed to be periodic in $\hat{x}$ with a periodicity of $\Lambda_x = \beta \lambda = 2\pi c_0/\omega_0$. Thus, the fields can be expressed as a Fourier series.
\begin{equation}
	E_x(x,\omega) = E_0(\omega) \sum_{m=-\infty}^{\infty} T_m(\omega) e^{i m x \omega_0 / \beta c_0}
\end{equation}
where the $T_m(\omega)$ terms are the spatial Fourier amplitudes of the transfer function $H(x,\omega)$. See Ref.~\cite{plettner2009photonic} for a similar discussion.

The acceleration gradient at frequency $\omega$, $G(\omega)$, can be written as the average $E_x$ felt by the particle as it moves with velocity $\beta c_0 \hat{x}$ through the entire interaction length of the structure from $x= -L/2$ to $x=L/2$.
\begin{align}
G(\omega) &= \frac{1}{L}\int_{-L/2}^{L/2}dx\ E_x(x,\omega) e^{i x \omega / \beta c_0} \\
		 &= \frac{1}{L}\int_{-L/2}^{L/2}dx\ E_0(\omega) \sum_{m=-\infty}^{\infty} T_m(\omega) e^{i (m\omega_0+\omega) x / \beta c_0}. 
\end{align}
Rearranging the integral and defining the normalized gradient $g(\omega) \equiv G(\omega)/E_0(\omega)$,
\begin{align}
g(\omega) &= \frac{1}{L} \sum_{m=-\infty}^{\infty} T_m(\omega) \int_{-L/2}^{L/2}dx\ e^{i (m\omega_0+\omega) x / \beta c_0)}.\\ 
 &= \sum_{m=-\infty}^{\infty} T_m(\omega) \frac{2\beta c_0 \sin\Big(\frac{L}{2\beta c_0}(m\omega_0+\omega)\Big)}{L(m\omega_0+\omega)}\\
 &= \sum_{m=-\infty}^{\infty} T_m(\omega) \ 
 \rm{sinc} \Big( \frac{L}{2\beta c_0}(m\omega_0+\omega) \Big).
\end{align}
We reasonably assume that the input pulse power is centered around $\omega_0$. In this case, then only the $m=-1$ will contribute to the accelerating mode. We could have also chosen a higher order $m = -2,-3,..$ for the accelerating mode, as was demonstrated previously \cite{breuer2013laser, mcneur2016laser}, but $m=-1$ is chosen for simplicity. Thus, as the interaction length increases, the $\rm{sinc}()$ function becomes more tightly centered around $\omega = \omega_0$. This has the effect of limiting the available bandwidth of the input pulse.

Under this assumption, the final form of the normalized gradient becomes
\begin{equation}
 g(\omega) = T_{-1}(\omega) \ 
 \rm{sinc} \Big( \frac{L}{2\beta c_0}(\omega-\omega_0) \Big).
\end{equation}
Assuming $T_{-1}(\omega)$ is relatively constant over a bandwidth larger than our input pulse, then we see that the gradient falls to zero at $\omega = \omega_0 \pm \frac{2\pi\beta c_0}{L}$. For a Gaussian pulse of duration $\tau$ with a time-bandwidth product of 0.44, the gradient would fall to zero at 
\begin{equation}
L = \tau \frac{4\pi\beta c_0}{0.44}.
\label{eq:L_tau}
\end{equation}

For a $\tau$ of 250\,fs and $\beta$ of 1, this corresponds to an interaction length of 2.14 mm. Thus, in order to satisfy the bandwidth requirement, $L$ must be much less than 2.14\,mm if no pulse delay techniques are used. 

This can also be estimated an a simple fashion. An electron traveling over a length $L$ with speed $\beta c_0$ will spend $\Delta t_{e^-} = \frac{L}{\beta c_0}$ of time in the channel. The input pulse will spend approximately $\tau$ seconds in the gap. Thus, for the fields to be present during the whole duration
\begin{equation}
L < \tau \beta c_0,
\end{equation}
which scales with $\tau$, $\beta$, and $c_0$ in the same fashion as Eq.~\ref{eq:L_tau}.


\bibliography{apssamp}

\end{document}